\documentclass[12pt]{iopart}

\begin{document}

\title[Non-Markovian master equation for a damped driven two-state system]{Non-Markovian master equation for a damped driven two-state system}

\author{P Haikka}

\address{Department of Physics and Astronomy, University of Turku, 20014 Turku, Finland}
\ead{pmehai@utu.fi}

\begin{abstract}
We present a detailed microscopic derivation for a non-Markovian master equation for a driven two-state system interacting with a general structured reservoir. The master equation is derived using the time-convolutionless projection operator technique in the limit of weak coupling between the two-state quantum system and its environment. We briefly discuss the Markov approximation, the secular approximation and their validity.
\end{abstract}

%Keywords required only for MST, PB, PMB, PM, JOA, JOB? 
 %Uncomment for Submitted to journal title message
 %Comment out if separate title page not required
\maketitle

\section{Introduction}
All quantum systems are open, i.e., they interact with their environment. This leads to dissipation and to a loss of quantum coherence due to an irreversible flow of energy and/or information from the system to its environment \cite{breuer&petruccione, weiss}. In some cases, however, it is possible that the properties of the environment modify the dynamics of dissipation and decoherence. In particular, reservoir memory effects can damp or even momentarily reverse the direction of the flow of energy and information, and the quantum system can regain some information that was previously lost to the environment \cite{nmqj, nmqj2}.\\
The properties of the environment are described by the spectral density function $J(\omega)$. When the spectral density is structured, i.e., $J(\omega)$ varies considerably with frequency, environmental memory effects can become significant. Physical examples of environments with a structured spectral density include photonic band-gap materials and leaky optical cavities \cite{haroche, lambropoulos}.\\
When the reservoir memory effects considerably modify the dynamics of the quantum system, the dynamical semigroup property is violated and a non-Markovian description of the dynamics is called for \cite{dynamical semigroups}. A systematic method for deriving a non-Markovian master equation is given by projection operator techniques \cite{nakajima, zwanzig}. In general, a non-Markovian master equation contains an integration over the whole past history of the system of interest. However, in the limit of weak coupling between the system and the environment one may use the time-convolutionless (TCL) projection operator technique to obtain a simpler local-in-time non-Markovian master equation \cite{shibata}.\\
In this paper we derive the non-Markovian master equation for a damped driven two-state system in the limit of weak coupling between the quantum system of interest and its environment. We assume the environment to be described by a generic structured spectral density.  A microscopic master equation for the damped driven two state system in the Markovian regime has been derived by Tana\'s and Kowalewska-Kud\l aszyk for the special case of a Lorentzian environment \cite{tanas}. The non-Markovian dynamics of such a system has been studied by Budini, using a phenomenological master equation \cite{budini}. We present a microscopic derivation for a non-Markovian master equation. The microscopic master equation is useful when we wish to understand the microscopic processes that determine the dynamics of the quantum system. 

\section{Derivation of the non-Markovian master equation}
\subsection{The model}
We consider a two-level atom with an energy separation $\omega_A$  ($\hbar=1$) driven by a monochromatic laser of frequency $\omega_L$ almost resonant with the frequency of the atom $|\omega_A-\omega_L|\ll\omega_A$. The two-level atom interacts with a quantised electromagnetic field modelled by an infinite chain of quantum harmonic oscillators of frequencies $\{\omega_k\}$. For convenience we study the dynamics of such system in a frame rotating at the laser frequency $\omega_L$.  In the rotating frame the free Hamiltonians of the system of interest and of the environment, and the interaction Hamiltonian are
\begin{eqnarray}
H_S=\frac{1}{2}(\Delta\sigma_z+\Omega\sigma_x), \label{hs}\\
H_E=\sum_k\omega_k a_k a_k^{\dagger},\\
H_I=\sum_k g_k[\exp(\rmi\omega_L t)\sigma_+a_k+\exp(-\rmi\omega_L t)\sigma_-a_k^{\dagger}],\label{hi}
\end{eqnarray}
respectively. Here $\sigma_{x,y,z}$ are the usual Pauli operators and $\sigma_{\pm}$ are the raising and lowering operators. The operators are given in the atomic basis, i.e., an orthonormal basis $\{|e\rangle, |g\rangle\}$ determined by the excited and ground state of the two-level atom. $\Omega$ is the Rabi frequency of the laser and $g_k$ is a coupling term describing the interaction between the two-level atom and the $k$-th quantum harmonic oscillator. We also defined the detuning $\Delta=\omega_A-\omega_L$.\\
The system Hamiltonian (\ref{hs}) and the interaction Hamiltonian (\ref{hi}) are both given in the rotating wave approximation, i.e., we have neglected non-energy conserving terms. 

\subsection{The TCL2 master equation}
According to the time-convolutionless projection operator technique a general master equation to second order in coupling between the system of interest and its environment is
\begin{equation} \label{me}
\frac{\rmd }{\rmd t}\rho(t)=-\int_0^tds\,\tr_E\Big[H_I(t),[H_I(s),\rho(t)\otimes\rho_E]\Big],
\end{equation}
where 
\begin{equation}
\label{ }
H_I(t)=\exp[\rmi t(H_S+H_E)]H_I\exp[-\rmi t(H_S+H_E)]
\end{equation}
is the interaction Hamiltonian in the interaction picture. We have assumed factorized initial conditions for the total density matrix $\rho_{tot}(0)=\rho(0)\otimes\rho_E(0)$. Furthermore, in the limit of weak coupling, the state of the environment remains stationary and therefore $\rho_E(t)=\rho_E(0)\equiv\rho_E$ for all times $t$.

\subsection{Transformation to the eigenbasis}
For convenience we go to the eigenbasis of the free system Hamiltonian $H_S$, i.e., orthonormal basis defined by 
\begin{equation}
|\psi_{\pm}\rangle=\pm\frac{1}{\sqrt{2}}\left(\sqrt{1\pm\sin\theta}|e\rangle+\sqrt{1\mp\sin\theta}|g\rangle\right),
\end{equation}
where $\theta=\arctan(\Delta/\Omega).$ In the eigenbasis representation the system Hamiltonian is diagonal, $H_S=(\omega/2)\bar{\sigma}_z,$ where $\omega=\sqrt{\Delta^2+\omega^2}$ is the energy bias between the eigenstates and a bar over an operator indicates that the operator is given in the eigenbasis. The interaction Hamiltonian in the interaction picture becomes
\begin{equation}
\label{int}
H_I(t)=\sum_{i=1,2}S_i(t)\otimes E_i(t),
\end{equation}
where
\begin{eqnarray}
\label{s}
S_1(t)=\frac{\exp(\rmi \omega_L t)}{2\omega}\left[\exp(-\rmi \omega t)(\Delta-\omega)\bar{\sigma}_-+\exp(\rmi \omega t)(\Delta+\omega)\bar{\sigma}_++\Omega\bar{\sigma}_z\right],\\
E_1(t)=\sum_k\exp(-\rmi \omega_k t)g_ka_k, \label{e}
\end{eqnarray}
and $S_2(t)=S_1^{\dagger}(t)$ and $E_2(t)=E_1^{\dagger}(t)$. 

\subsection{The environment}
We assume that the environment is in a thermal state $\rho_E=Z^{-1}\exp\left(-H_E/k_B T\right)$, where $Z$ is the partition function and $k_B$ is the Boltzmann constant. We will consider the case when temperature $T=0$ and only a single reservoir correlation function is non-zero:
\begin{eqnarray}
\label{ }
\tr_E[E_i(t_2)E_j(t_1)\rho_E]=0,\qquad i\neq2, j\neq1,\nonumber\\
\tr_E[E_2(t_2)E_1(t_1)\rho_E]=\sum_k\exp [\rmi \omega_k (t_2-t_1)]g_k^2. \label{correlation}
\end{eqnarray}
In the limit of a continuum of modes of the environment
 \begin{equation}
\label{limit}
\sum_kg_k^{2}\rightarrow\int\rmd\omega\,J(\omega),
\end{equation}
where we introduce the generic spectral density $J(\omega)$. Combining equations (\ref{me}), (\ref{int}), (\ref{correlation}) and (\ref{limit}) we find
\begin{eqnarray}
\label{apu}
\frac{\rmd }{\rmd t}\bar{\rho}(t)&=&\int_0^t\rmd\tau\int\rmd\omega'\,J(\omega')\rme^{-\rmi\omega't}[S_2(t-\tau)\bar{\rho}(t)S_1(t)\nonumber\\
&-&S_1(t)S_2(t-\tau)\bar{\rho}(t)]+H.c.,
\end{eqnarray} 
where $H.c.$ is an abbreviation for Hermitian conjugation. At this point we define
\begin{equation}
\label{G}
\Gamma_{\xi}(t)=\int_0^t\rmd\tau\int\rmd\omega'\,J(\omega')\exp[\rmi(\omega_L-\omega'+\xi\omega)t],\qquad\xi\in\{-1,0,1\}.
\end{equation}
Recall that the detuning is small, i.e., $|\Delta|\ll\omega_A,\omega_L$. Furthermore, in physically realistic situations the Rabi frequency is also small, i.e. $\Omega\ll\omega_A,\omega_L$. Therefore $\omega=\sqrt{\Delta^2+\Omega^2}\ll\omega_L$ and $\Gamma_{\pm1}(t)\approx\Gamma_0(t)\equiv\Gamma(t)$. In the following we decompose the quantity $\Gamma(t)$ into real and imaginary parts, i.e. $\Gamma(t)=(1/2)\gamma(t)+\rmi\lambda(t)$.\\ 
Keeping this in mind we obtain form equation (\ref{apu}) the master equation for the damped driven two-state system in the Schr\"odinger picture
\begin{eqnarray} \label{master}
 \frac{\rmd }{\rmd t}\bar{\rho}(t)&=&-\rmi[H_S+H_L]+C_+^2\gamma(t)\left[\bar{\sigma}_-\bar{\rho}(t)\bar{\sigma}_+-\frac{1}{2}\{\bar{\sigma}_+\bar{\sigma}_-,\bar{\rho}(t)\}\right]\nonumber\\
&+&C_-^2\gamma(t)\left[\bar{\sigma}_+\bar{\rho}(t)\bar{\sigma}_--\frac{1}{2}\{\bar{\sigma}_-\bar{\sigma}_+,\bar{\rho}(t)\}\right]\nonumber\\
&+&C_0^2\gamma(t)\left[\bar{\sigma}_z\bar{\rho}(t)\bar{\sigma}_z-\frac{1}{2}\{\bar{\sigma}_z\bar{\sigma}_z,\bar{\rho}(t)\}\right]\nonumber\\
&-&C_-C_0\gamma(t)\left[\bar{\sigma}_+\bar{\rho}(t)\bar{\sigma}_z+\bar{\sigma}_z\bar{\rho}(t)\bar{\sigma}_-\right]\nonumber\\
&+&C_+C_0\gamma(t)\left[\bar{\sigma}_-\bar{\rho}(t)\bar{\sigma}_z+\bar{\sigma}_z\bar{\rho}(t)\bar{\sigma}_+\right]\nonumber\\
&+&C_0^2\gamma(t)[\bar{\sigma}_+\bar{\rho}(t)\bar{\sigma}_++\bar{\sigma}_-\bar{\rho}(t)\bar{\sigma}_-]\nonumber\\
&+&C_0\left[\frac{\gamma(t)}{2}\{\bar{\sigma}_x,\bar{\rho}(t)\}+\rmi\lambda(t)[\bar{\sigma}_x,\bar{\rho}(t)]\right],
\end{eqnarray}
where the Lamb shift Hamiltonian is
\begin{equation}
\label{ }
H_L=\lambda(t)[C_+^2\bar{\sigma}_-\bar{\sigma}_++C_+^2\bar{\sigma}_-\bar{\sigma}_-+C_0^2\bar{\sigma}_z^2],
\end{equation}
and 
\begin{equation}
\label{ }
C_\pm=\frac{\omega\pm\Delta}{2\omega},\qquad C_0=\frac{\Omega}{2\omega}.
\end{equation}

\subsection{Markov approximation}
The reservoir correlation time $\tau_C$ is defined by the properties of the environment. If $\tau_C$ is small compared to the relaxation time of the system $\tau_R$, the reservoir memory effects are negligible and one can make the Markov approximation. When the Markov approximation is valid one can extend the upper limit of integration in equation (\ref{G}) from $t$ to infinity. In the Markovian limit the decay rates $\gamma(t)$ become time independent positive constants $\gamma=\lim_{t\rightarrow0}\gamma(t)$.\\
We note that in the limit of weak approximation between the system of interest and its environment the first non-Markovian corrections do not change the operatorial form of the Markovian master equation. Instead the first non-Markovian corrections only affect the decay rates by making them time-independent. For times $t=\mathcal{O}(\tau_C)$ the time-dependent decay rates deviate from their stationary Markovian values and it is during this timescale that non-Markovian reservoir memory effects take place.

\subsection{The secular approximation}
The typical timescale for the damped driven two-state system is $\tau_S=\omega^{-1}$. When the typical timescale is much smaller than the relaxation time $\tau_R$ one can make the secular approximation (rotating wave approximation) on the master equation. The secular approximation comprises of neglecting terms that, in the interaction picture, oscillate rapidly with frequency $\tau_S^{-1}$. In the case considered in this paper the secular approximation amounts to neglecting the last four lines of the master equation (\ref{master}).\\
When the secular approximation is valid the master equation consists of a coherent part describing unitary evolution generated by $H_S+H_L$, and of a part with three terms in a Lindblad-type form, i.e., in a Lindblad form with time-dependent coefficients. The Lindblad-type terms describe quantum jumps that disrupt the unitary evolution of the system.\\
When $\gamma(t)\geq0$ one can describe the quantum jumps in the framework of Monte Carlo wave functions \cite{mcwf}. In this case the quantum jumps lead to irreversible dissipation and decoherence of the quantum system. If, however, $\gamma(t)<0$ for some period of time, a more general description, namely the non-Markovian quantum jump (NMQJ) method is called for \cite{nmqj}. For time interval when the decay rate is negative the direction of the quantum jumps is reversed. Reversed quantum jumps reverse dissipative processes and allow the quantum system to regain some of the information and/or energy that was previously lost to the environment. The system can recohere and reconstruct superposition states. The NMQJ description explains how memory effects affects the dynamics of a quantum system described by a local-in-time non-Markovian master equation. 

\section{Conclusions}
We have derived a local-in-time non-Markovian master equation for a damped driven two-state system interacting with a general structured environment using the time-convolutionless projection operator technique to second order in coupling between the system and its environment. We assumed factorized initial conditions and a stationary reservoir. To keep master equation (\ref{master}) as general as possible we did not perform the secular approximation.\\ 
We found that to second order in perturbation theory the operatorial form of the non-Markovian master equation is the same as the operatorial form of the corresponding Markovian master equation. The important difference that arises from the first non-Markovian corrections is that, for timescales specified by the reservoir memory time, the decay rates become time-dependent.\\
The specific form of the decay rates can be computed once the environment specific spectral density function has been specified. Master equation (\ref{master}) lays the basis for a systematic study of non-Markovian dynamics of the damped driven two-state system embedded in different types of reservoirs. 

\ack
The author acknowledges financial support from the Eemil Aaltonen foundation, and thanks Sabrina Maniscalso for many useful discussions and for her comments on this paper.

\end{document}